\documentclass[prl,twocolumn,secnumarabic,superscriptaddress,showpacs,amssymb,amsmath,nobibnotes,aps]{revtex4}

\usepackage{latexsym}
\usepackage{amsthm,amsfonts,url}   

\newtheorem{theorem}{Theorem}
\newtheorem{definition}{Definition}
\newtheorem{Cor}{Corollary}
\newtheorem{lem}{Lemma}

\begin{document}

\title{Local Distinguishability of Any Three Quantum States}

\author{Somshubhro Bandyopadhyay}
\email{bandyo@iro.umontreal.ca}
\affiliation{Institute for Quantum Information Science, University of Calgary, Alberta T2N 1N4, Canada.}
\affiliation{DIRO, Universit\'e de Montr\'eal, C.\,P.~6128, Succursale Centre-Ville, Montr\'eal, Qu\'ebec, Canada H3C 3J7.}

\author{Jonathan Walgate}
\email{jwalgate@perimeterinstitute.ca}
\affiliation{Institute for Quantum Information Science, University of Calgary, Alberta T2N 1N4, Canada.}
\affiliation{Perimeter Institute for Theoretical Physics, 31 Caroline St. N, Waterloo, Ontario, Canada N2L 2Y5.}

\date{\today}

\begin{abstract}
We prove that any three linearly independent pure quantum states can always be locally distinguished with nonzero probability regardless of their dimension, entanglement, or multipartite structure. Almost always, all three states can be unambiguously identified. The only exceptional case, where one state is locally knowable but the other two are not, is found among multi-qubit states.
\end{abstract}


\maketitle

Global operations on a quantum system can process information in ways that local operations on the system's parts cannot. All
uses of entanglement in quantum information theory flow from this one fact, from teleportation~\cite{teleport} to Shor's
factoring algorithm~\cite{Shor}. However a fundamental question remain unanswered. When is global information about a quantum
system also available locally? This question can be formally posed as a local state discrimination task. Given one copy of a
system in one of a known set of quantum states $\{ |\psi_{i}\rangle \}$, how much `which state' information can be gleaned by
local operations and classical communication (LOCC), and how much more information is revealed by global measurements?

This problem has attracted much attention in recent years, after surprising results showed perfect local distinguishability was
not directly linked to entanglement. Bennett and coworkers presented sets of orthogonal unentangled states that were not
perfectly locally distinguishable~\cite{sausages}. JW, Short, Hardy and Vedral proved orthogonal pairs of states are always
perfectly locally distinguishable, irrespective of their entanglement~\cite{Walgate1}.

There are two natural approaches to quantum state discrimination.
Optimal discrimination seeks the best possible guess as to the state
of the system~\cite{optimal}. Conclusive discrimination (also called
unambiguous discrimination) seeks certain knowledge of the state of
the system, balanced against a possibility of
failure~\cite{conclusive}. It follows directly from the results of
Walgate \emph{et al.}~\cite{Walgate1} and Virmani \emph{et
al}.~\cite{Virmani} that local parties can always gain some amount
of `which state' information about the pure state of a shared
system, and use it to improve their guesswork. Optimal state
discrimination is always locally feasible in this sense, although
the local optimum may be significantly worse than the global.
Conclusive discrimination is more interesting. All pairs of pure
quantum states can be conclusively discriminated equally well
locally and globally~\cite{Ji}. Generically, a small number of pure
states (proportional to the dimension of the subsystems) can be
conclusively discriminated with nonnegligable probability~\cite{WS}.
But in every multipartite dimensionality there are sets of four pure
quantum states that are not conclusively locally distinguishable
\emph{at all}. In this case local parties can never gain certain
knowledge of which state they possess; the Bell states are the
simplest example of such a set~\cite{Ghosh}.

So two states are always conclusively distinguishable, and four states can be conclusively indistinguishable. We complete the picture by showing that provided they are linearly independent (only linearly independent states are
globally distinguishable) three pure quantum states can be conclusively locally distinguished. Local protocols may not succeed as
often as global measurements, but they can succeed some of the time. No triplet of pure states, no matter how entangled, conceals
any fraction of its `which state' information from local parties with certainty.
\bigskip

We present our results in the following framework. A multipartite quantum system $Q$ is shared between $n$ different local
parties, each with access to one of $n$ local Hilbert spaces: $\mathcal{H}_{Q} = \bigotimes_{j=1}^{n} \mathcal{H}_{j}$. It has
been prepared in one of a known set of possible pure states $\mathcal{S} = \{|\psi_{i}\rangle \}$, each with some nonzero (but
potentially unknown) probability $p_{i}$. The local parties are set the task of discovering with certainty which of the states
$\mathcal{S}$ they have been given, using only LOCC. We will use the following definitions.
\begin{definition}
A state $|\psi_{i}\rangle \in \mathcal{S}$ is \textbf{conclusively locally identifiable} if and only if there is a LOCC protocol
whereby with some nonzero probability $p>0$ it can be determined that $Q$ was certainly prepared in state $|\psi_{i}\rangle$.
\end{definition}
\begin{definition}
A set of states $\mathcal{S}$ is \textbf{conclusively locally distinguishable} if and only if every state in $\mathcal{S}$ is conclusively locally identifiable.
\end{definition}


Conclusive state identification has qualitative links to entanglement. It was proved by Horodecki \emph{et al.} that the states of a complete
orthonormal basis are conclusively locally identifiable states if and only if they are product states \cite{Horodecki}. We show below in Corollary~\ref{cor1} that if no members of an incomplete basis of orthogonal states are conclusively identifiable then the set must be completely entangled.

We begin by establishing a necessary and sufficient condition for a set of states to be conclusively locally distinguishable,
first proved by Chefles~\cite{Chefles}. We outline a simplified version of Chefles' specific to the case of pure states. We will
then show how this condition holds for sets of three states.

\begin{lem}[Chefles] \label{productstatestrategy}
Let a multipartite quantum system $Q$ be prepared in one of a set of pure, linearly independent multipartite quantum states
$\mathcal{S} = \{|\psi_{i}\rangle \}$. Let $|\psi_{x}\rangle \in \mathcal{S}$.

If and only if there exists a product state $|\phi\rangle$ such that $\forall i\neq x
\ \langle \psi_{i} | \phi \rangle = 0$, and $\langle \psi_{x} | \phi \rangle \neq 0$, then $|\psi_{x}\rangle$ is conclusively
locally identifiable in $\mathcal{S}$.
\end{lem}

Proof of sufficiency: Assume a products state $|\phi\rangle$ with the above properties exists. The parties can locally project into a product
basis of $\mathcal{H}_{Q}$ that includes $|\phi\rangle$. If the state of $Q$ is $|\psi_{x}\rangle$ they will obtain the result
projecting onto $|\phi\rangle$ with probability $|\langle \psi_{x} |\phi\rangle|^{2}$, which is greater than zero. In this case,
they have conclusively locally identified $|\psi_{x}\rangle$ since no other state $|\psi_{i}\rangle$ ever yields this projection
result. $\Box$

Proof of necessity: Assume that $|\psi_{x}\rangle$ is conclusively
locally identifiable. There is a LOCC protocol, describable by a separable superoperator,
which can produce at least one measurement outcome conclusively identifying $|\psi_{x}\rangle$. This outcome corresponds to some
separable POVM element $M^{\dagger}M = A^{\dagger}A \otimes B^{\dagger}B \otimes ...$, which because it identifies
$|\psi_{x}\rangle$ must satisfy $\forall i\neq x \ \langle \psi_{i} | M^{\dagger}M |\psi_{i}\rangle = 0$, and $\langle \psi_{x} |
M^{\dagger}M |\psi_{x}\rangle \neq 0$. $M^{\dagger}M$ is decomposable into a set of rank one projection operators onto product
states $\{ |P_{l}\rangle \}$: $M^{\dagger}M = \sum_{jk...} A_{j}^{\dagger}A_{j} \otimes B_{k}^{\dagger}B_{k} \otimes ... =
\sum_{l} (|P_{l}\rangle\langle P_{l}|)^{\dagger}(|P_{l}\rangle\langle P_{l}|)$. These product states must satisfy $\forall l \forall i\neq x \ \langle \psi_{i} | P_{l}^{\dagger}P_{l} |\psi_{i}\rangle = 0$, and $\exists l \ \langle \psi_{x} | P_{l}^{\dagger}P_{l} |\psi_{x}\rangle \neq 0$. Let the product state satisfying both conditions be $|\phi\rangle$. Thus there exists a product state $|\phi\rangle$ such that $\forall i\neq x \ \langle \psi_{i} | \phi \rangle = 0$, and $\langle \psi_{x} | \phi \rangle \neq 0$. $\Box$

\begin{Cor}\label{cor1}
All product states belonging to sets of pure orthogonal states are conclusively locally identifiable, and the subset of unentangled members
of such a set is conclusively locally distinguishable.
\end{Cor}

Proof: If $\mathcal{S} = \{|\psi_{i}\rangle \}$ is a set of orthogonal pure states, and $|\psi_{x}\rangle \in \mathcal{S}$ is a product state, then $| \phi \rangle = |\psi_{x}\rangle $ satisfies the sufficient condition of Lemma~\ref{productstatestrategy}. $\Box$
\bigskip

Interestingly, although sets of pure orthogonal states must be completely entangled in order to be completely conclusively indistinguishable,
linearly independent states are not so restricted. In fact, Duan \emph{et al.} have recently shown that there are sets of product
states that are completely conclusively indistinguishable - a strong form of `nonlocality without entanglement'~\cite{Duan}. JW and Scott have shown that generic sets of states obey the same numerical threshold for conclusive distinguishability whether they are entangled or not~\cite{WS}.

\begin{theorem} \label{theorem1}
Let a multipartite quantum system $Q$ be prepared in one of a set of three pure, linearly independent multipartite quantum states $\mathcal{S} = \{|\psi_{1}\rangle, |\psi_{2}\rangle, |\psi_{3}\rangle \}$.

There exists $ |\psi_{x}\rangle \in \mathcal{S}$ such $ |\psi_{x}\rangle$ is conclusively locally identifiable.
\end{theorem}

We will prove this result separately for three different cases. First we will deal with systems whose three possible states cannot be composed on a chain of qubits (i.e. where $\text{Span} (\mathcal{S}) \subset \mathcal{H}_{Q} \neq \bigotimes_{i=1}^{n} \mathcal{H}_{2}$), with each local party holding just one qubit). Then we will consider $\mathcal{H}_{2} \otimes \mathcal{H}_{2}$ systems. Lastly, we will prove our result for larger arrays of qubits: $\mathcal{H}_{Q} = \bigotimes_{i=1}^{n>2} \mathcal{H}_{2}$. These three cases cover all possible multipartite situations.

\begin{lem}[Higher-dimensional states] \label{lemma2}
Let a multipartite quantum system $Q$ be prepared in one of a set of three pure, linearly independent multipartite quantum states $\mathcal{S} = \{|\psi_{1}\rangle, |\psi_{2}\rangle, |\psi_{3}\rangle \}$. Let the space spanned by $\mathcal{S}$ be such that it cannot be expressed in the form $\bigotimes_{i=1}^{n} \mathcal{H}_{2}$.

$\mathcal{S}$ is conclusively locally distinguishable.
\end{lem}

Proof: If the space spanned by $\mathcal{S}$ cannot be expressed in the form $\bigotimes_{i=1}^{n} \mathcal{H}_{2}$, then at least one of the local parties has an irreducibly three- or higher-dimensional Hilbert space $\mathcal{H}_{i}$. We call this party `Alice'. We can write the states thus:
\begin{align}
|\psi_{1} \rangle &= \sum_{i} a_{i} |i\rangle_{A} |\eta_{i}\rangle_{BC...} , \nonumber\\
|\psi_{2} \rangle &= \sum_{i} b_{i} |i\rangle_{A} |\nu_{i}\rangle_{BC...} , \label{form}\\
|\psi_{3} \rangle &= \sum_{i} c_{i} |i\rangle_{A} |\mu_{i}\rangle_{BC...}, \nonumber
\end{align}
where the vectors $|\eta_{i}\rangle_{BC...}, |\nu_{i}\rangle_{BC...}$, and $|\mu_{i}\rangle_{BC...}$ are normalized, and $a_{i}, b_{i}$ and $c_{i}$ are complex coefficients satisfying $\sum_{i} a_{i}^{*}a_{i} = 1$. Following the strategy of Lemma~\ref{productstatestrategy}, we will show that there exists a product state $|\phi\rangle$ such that $\langle \psi_{1} | \phi \rangle = \langle \psi_{2} | \phi \rangle = 0$ and $\langle \psi_{3} | \phi \rangle \neq 0$. Let us write the product state thus:
\begin{displaymath}
|\phi \rangle = ( \sum_{i} x_{i} |i\rangle_{A} ) \otimes | \theta \rangle_{BC...},
\end{displaymath}
with $\sum_{i} x_{i}^{*}x_{i} = 1$. We choose $| \theta \rangle$ such that it is a product state amongst the parties $B,C...$ and so that the equations (\label{threeeqns}) below are linearly independent. We can always do this. (Note in fact that a randomly chosen $| \theta \rangle$ will have this property with probability one, thanks to the linear independence of the states in $\mathcal{S}$.) $|\phi\rangle$ must satisfy the following conditions:
\begin{align}
\langle \psi_{1} |\phi \rangle &=  \sum_{i} x_{i} \ a_{i}^{*} \langle \eta_{i} | \theta \rangle = 0, \nonumber \\
\langle \psi_{2} |\phi \rangle &=  \sum_{i} x_{i} \ b_{i}^{*} \langle \nu_{i} | \theta \rangle = 0, \label{threeeqns} \\
\langle \psi_{3} |\phi \rangle &=  \sum_{i} x_{i} \ c_{i}^{*} \langle \mu_{i} | \theta \rangle \neq 0. \nonumber
\end{align}
The quantities $a_{i}^{*} \langle \eta_{i} | \theta \rangle$,
$b_{i}^{*} \langle \nu_{i} | \theta \rangle$, and $c_{i}^{*} \langle
\mu_{i} | \theta \rangle$ are all fixed by our arbitrary choice of
basis $\{ |i\rangle_{A} \}$, and product state
$|\theta\rangle_{BC...}$. There are at least three variables
$x_{i}$, because $\mathcal{H}_{A} \neq 2$. With three linearly
independent equations and three variables, there is always a
solution for the $x_{i}$. (Note that normalization does not further
restrict the solution of these equations, as they only specify sums
to `zero' or `not zero'.) Therefore, we can always find a product
state $|\phi \rangle$ that is orthogonal to $|\psi_{1} \rangle$ and
$|\psi_{2} \rangle$, but nonorthogonal to $|\psi_{3} \rangle$. By
Lemma~\ref{productstatestrategy}, this means $|\psi_{3} \rangle$ is
conclusively locally identifiable in $\mathcal{S}$.

The same reasoning applies to $|\psi_{1} \rangle$ and $|\psi_{2} \rangle$, so $\mathcal{S}$ is conclusively locally distinguishable. $\Box$

Surprisingly, the only exceptions to this `one for all and all for one' structure are found amongst the simplest quantum systems - qubits.


\begin{lem}[Two Qubits] \label{lemma3}
Let a multipartite quantum system $Q$ be prepared in one of a set of three pure, linearly independent multipartite quantum states $\mathcal{S} = \{|\psi_{1}\rangle, |\psi_{2}\rangle, |\psi_{3}\rangle \}$. Let $\mathcal{H}_{Q} = \mathcal{H}_{2} \otimes \mathcal{H}_{2}$.

There exists $ |\psi_{x}\rangle \in \mathcal{S}$ such $ |\psi_{x}\rangle$ is conclusively locally identifiable.
\end{lem}

Proof: Either at least two of the three members of $\mathcal{S}$ are product states, or else at least two of them are entangled states. Whichever is the case, we label the states such that $|\psi_{1} \rangle$ and $|\psi_{2} \rangle$ are similar - either they're both product states, or they're both entangled. We will show that $|\psi_{3} \rangle$ is then conclusively locally identifiable. $|\psi_{1} \rangle$ and $|\psi_{2} \rangle$ are linearly independent and span a two-dimensional subspace of $\mathcal{H}_{Q}$. Let us call this subspace ${H}_{a}$, and its complementary subspace ${H}_{a}^{\perp}$.


If $|\psi_{1} \rangle$ and $|\psi_{2} \rangle$ are product states, it follows trivially from their linear independence that ${H}_{a}$ can be spanned by a pair of orthogonal product states. Its complementary subspace ${H}_{a}^{\perp}$ must also be spanned by a pair of orthogonal product states.

If $|\psi_{1} \rangle$ and $|\psi_{2} \rangle$ are entangled states, exactly the same is true - both ${H}_{a}$ and ${H}_{a}^{\perp}$ must be spanned by a pair of product states. This is easy to see directly - writing the states in the general form $|\psi_{1} \rangle= a|00\rangle + b|11\rangle$ and $|\psi_{2} \rangle= c|01\rangle + d|10\rangle$, where $a,b,c$ and $d$ are nonzero complex numbers, the states satisfying $|\phi_{A} \rangle |\phi_{B} \rangle = |\psi_{1} \rangle + \sqrt{\frac{ab}{cd}} |\psi_{2} \rangle $ are the (unnormalised) product states spanning ${H}_{a}$. ${H}_{a}^{\perp}$ is spanned by $\{ |\phi_{A} \rangle |\phi_{B}^{\perp} \rangle \ , \ |\phi_{A}^{\perp} \rangle |\phi_{B} \rangle)$.

$|\psi_{3} \rangle$ is linearly independent of $|\psi_{1} \rangle$ and $|\psi_{2} \rangle$, so it has at least some support on ${H}_{a}^{\perp}$. It therefore has at least some support on one of the two product states spanning ${H}_{a}^{\perp}$. Let this product state be $|\phi\rangle$. In line with Lemma~\ref{productstatestrategy}, $|\phi \rangle$ is orthogonal to $|\psi_{1} \rangle$ and $|\psi_{2} \rangle$, but nonorthogonal to $|\psi_{3} \rangle$, and therefore $|\psi_{3} \rangle$ is conclusively locally identifiable. $\Box$
\bigskip

By symmetry, if all the states in $\mathcal{S}$ are entangled, or if they are all product states, then they are all conclusively locally identifiable and $\mathcal{S}$ is conclusively locally distinguishable. If two of them are product states, it is again simple to show they are all conclusively locally identifiable (a consequence of the fact that every set of three orthogonal $2 \otimes 2$ states two of which are product states is perfectly locally distinguishable~\cite{Walgate2}). But an exception occurs when two of the states are entangled: only the product state can be conclusively locally identified. For example, the set of states:
\begin{align}
|\psi_{1} \rangle &= \alpha_{1}|0\rangle_{A}|0\rangle_{B} + \alpha_{2}|1\rangle_{A}|1\rangle_{B}, \nonumber \\
|\psi_{2} \rangle &= \beta_{1}|0\rangle_{A}|0\rangle_{B} + \beta_{2}|1\rangle_{A}|1\rangle_{B} , \label{2qubit}\\
|\psi_{3} \rangle &= |0\rangle_{A}|1\rangle_{B}, \nonumber
\end{align}
is not conclusively distinguishable. $|\psi_{3} \rangle$ is conclusively locally identifiable, but neither $|\psi_{1}\rangle$ nor $|\psi_{2}\rangle$ can satisfy the necessary condition for conclusive local identifiability established by Lemma~\ref{productstatestrategy}. This asymmetric possibility is unique to triplets of qubit states, but at least one state can always be identified.

\begin{lem}[Many Qubits] \label{lemma4}
Let a multipartite quantum system $Q$ be prepared in one of a set of three pure, linearly independent multipartite quantum states $\mathcal{S} = \{|\psi_{1}\rangle, |\psi_{2}\rangle, |\psi_{3}\rangle \}$. Let $\mathcal{H}_{Q} = \bigotimes_{i=1}^{n>2} \mathcal{H}_{2}$.

There exists $ |\psi_{x}\rangle \in \mathcal{S}$ such $ |\psi_{x}\rangle$ is conclusively locally identifiable.
\end{lem}

Proof: We begin by considering the Hilbert space of the system with Alice and Bob's subspaces combined into one four-dimensional subspace $\mathcal{H}_{AB}$. We can write the states thus:
\begin{align}
|\psi_{1} \rangle &=  \sum_{i,j...=1}^{2} a_{ij...}  |\eta_{ij...}\rangle_{AB} |ij...\rangle_{CD...}, \nonumber \\
|\psi_{2} \rangle &=  \sum_{i,j...=1}^{2} b_{ij...}  |\nu_{ij...}\rangle_{AB} |ij...\rangle_{CD...}, \label{manyqubit}\\
|\psi_{3} \rangle &=  \sum_{i,j...=1}^{2} c_{ij...}  |\mu_{ij...}\rangle_{AB} |ij...\rangle_{CD...}. \nonumber
\end{align}
There are $n-2$ indices $i,j...$ . The states $\{ |ij...\rangle_{CD...} \}$ form an arbitrary canonical basis for the $\bigotimes^{n-2} \mathcal{H}_{2}$ Hilbert space shared by Carol, Douglas \emph{et al}. The complex coefficients $a_{ij...}, b_{ij...}$ and $c_{ij...}$ satisfy normalization constraints.

From Lemma~\ref{lemma2}, we know that a state $|\phi\rangle$ exists
that is unentangled under the $\mathcal{H}_{AB} \bigotimes^{n-2}
\mathcal{H}_{2}$ partition and which is orthogonal to $|\psi_{1}
\rangle$ and $|\psi_{2} \rangle$ but nonorthogonal to $|\psi_{3}
\rangle$. Let us write this state $|\phi\rangle = | \theta
\rangle_{AB} \otimes | \omega \rangle_{CD...}$. Our choice of
canonical basis $\{|ij...\rangle_{CD...}\}$ for equations
\ref{manyqubit} was arbitrary, so we can specify retroactively that
$| \omega \rangle_{CD...} = | 00...0\rangle_{CD...}$. Then we know
that $|\phi\rangle$ satisfies the following equations:
\begin{align}
\langle \psi_{1} |\phi \rangle &=  a_{ij...}^{*} \langle \eta_{ij...} | \theta \rangle = 0, \nonumber\\
\langle \psi_{2} |\phi \rangle &=  b_{ij...}^{*} \langle \nu_{ij...} | \theta \rangle = 0, \\
\langle \psi_{3} |\phi \rangle &=  c_{ij...}^{*} \langle \mu_{ij...} | \theta \rangle \neq 0. \nonumber
\end{align}
Clearly this can be true only if $|\mu_{ij...}\rangle_{AB}$ is linearly independent from both $|\eta_{ij...}\rangle_{AB}$ and $|\nu_{ij...}\rangle_{AB}$.

$|\eta_{ij...}\rangle_{AB}$ and $|\nu_{ij...}\rangle_{AB}$ are either linearly independent of one another, or they are identical. If they are identical, we can trivially find a candidate for $|\theta\rangle_{AB} $ that is a product state in $\mathcal{H}_{A} \otimes \mathcal{H}_{B}$, and $|\psi_{3}\rangle$ is conclusively identifiable state by Lemma~\ref{productstatestrategy}. If they are not identical, then $\{ |\eta_{ij...}\rangle_{AB}, |\nu_{ij...}\rangle_{AB}, |\mu_{ij...}\rangle_{AB} \}$ is a set of three pure linearly independent states, and from Lemma~\ref{lemma3} there is some product state $|\xi\rangle$ that is nonorthogonal to exactly one of them. In this case, the state $ |\xi\rangle_{AB} \otimes |\omega\rangle_{CD...}$ is a completely unentangled state in $\mathcal{H}_{Q}$ satisfying Lemma~\ref{productstatestrategy} for one of the three states in $\mathcal{S}$ (though not necessarily $|\psi_{3}\rangle$!). Therefore there is some state in $\mathcal{S}$ that is conclusively locally identifiable. $\Box$
\bigskip

This is the third and final step in our proof of Theorem~\ref{theorem1}. In all three possible cases, all triplets of pure linearly independent quantum states have been shown to contain a conclusively identifiable state.
\bigskip

If a set of states contains an identifiable member then complete `which state' information is potentially locally
discoverable. Otherwise it is necessarily hidden from local observation. In spite of the known links between entanglement and
conclusive identifiability, we have shown that any three states can always be locally induced to reveal this information with some probability, no matter how entangled. Furthermore, unless the triplet is a very specific set of multi-qubit states, it is conclusively locally distinguishable and all possible states can be unambiguously identified. An open question is finding optimal local protocols, which would allow a quantitative comparison of the local and global situation.

\section{Acknowledgements}

We would like to thank Runyao Duan, Aidan Roy, Anirban Roy, Barry Sanders and Andrew Scott for useful discussions. JW
acknowledges support from the Alberta Ingenuity Fund and the Pacific Institute for the Mathematical Sciences, and thanks DIRO at
the Universit\'e de Montr\'eal for their hospitality with assistance from the Canadian Institute for Advanced Research. SB
acknowledges support from Alberta's Informatics Circle of Research Excellence (iCORE), the Canadian Network of Centres of
Excellence for the Mathematics of Information Technology and Complex Systems (MITACS), the Canadian Institute for Advanced
Research, General Dynamics Canada, and the Natural Science and Engineering Research Council of Canada (NSERC).

\end{document}